\documentclass[pra,aps,superscriptaddress,showpacs]{revtex4-1}%,twocolumn

\usepackage[FIGTOPCAP,tight]{subfigure} %FIGTOPCAP
\usepackage{graphicx}% Include figure files
\usepackage{dcolumn}% Align table columns on decimal point
\usepackage{bm}% bold math

\usepackage[T1]{fontenc}
\usepackage[latin9]{inputenc}
\usepackage{geometry}
\usepackage{amssymb}
\usepackage{graphicx}
\usepackage{esint}
\usepackage{array}
\usepackage{multirow}
\usepackage{color} %for change tracking

\newcommand{\ket}[1]{\ensuremath{\left|#1\right\rangle}}

\begin{document}

\title{Nonlinear optics determination of the symmetry group of a crystal using structured light }

\author{Rocio J\'auregui}
\affiliation{Instituto de F\'{i}sica, Universidad Nacional
Aut\'{o}noma de M\'{e}xico, Apartado Postal 20-364, 01000
M\'{e}xico D.F., M\'{e}xico}

\email{rocio@fisica.unam.mx}

\author{Juan P. Torres}
\affiliation{ICFO---Institut de Ciencies Fot\'{o}niques,
Mediterranean Technology Park, 08860, Castelldefels, Barcelona,
Spain} \affiliation{Dep. Signal Theory and
Communications, Universitat Politecnica de Catalunya, Jordi Girona
1-3, 08034 Barcelona, Spain}

\begin{abstract}
We put forward a technique to unveil to which symmetry group a
nonlinear crystal belongs, making use of nonlinear optics with
structured light. We consider as example the process of
spontaneous parametric down-conversion. The crystal, which is
illuminated with a special type of Bessel beam, is characterized
by a nonlinear susceptibility tensor whose structure is dictated
by the symmetry group of the crystal. The observation of the
spatial angular dependence of the lower-frequency generated light
provides direct information about the symmetry group of the
crystal.
\end{abstract}

%42.70.Mp    Nonlinear optical crystals
%42.70.-a    Optical materials
%42.65.Lm    Parametric down conversion and production of entangled photons

\maketitle

%Introduction
The  spatial arrangements of atoms of substances that can
present themselves in a crystalline structure is
determined by the specific  ({\em point
group} or {\em symmetry group}) category \cite{sands} to which the
crystal belongs. According to Neumann's principle
\cite{neumann_principle}, or principle of symmetry, if a crystal
is invariant with respect to a set of symmetry transformations,
any physical property of the crystal is also invariant under such
operations. Thus,  the response of a crystal to
optical, electric or magnetic stimulus can be used to predict
general features of its structure, i.e., its symmetry group,
allowing to unveil its presence in a determined spatial region of
interest.

Crystallographic characterization  is usually performed by
diffraction techniques using X-rays\cite{x-ray}, or material
waves\cite{neutron-waves} \cite{electron-waves} that interact with
the atoms that are arranged in a certain pattern dictated by the
symmetry group. By observing the spatial distribution of the
photons or particles that come out from the crystal, one can
obtain the sought-after information. In spite of the high
importance of such techniques, sometimes they cannot be used, or
its use is cumbersome and extremely difficult. Moreover, they are
not immune to obstacles that may prevent a unique symmetry
assignment to a given diffraction pattern. This can be due, e. g.,
to insufficient number of Bragg peaks owing to a finite instrument
momentum range or spurious peaks arising from multiple scattering
events. In these cases, a companion technique might help to
identify the appropriate symmetry group of the crystal.

An alternative to diffraction techniques is the characterization
of the crystal symmetry by its effects on the nonlinear optical
response encoded in the susceptibility tensors
\cite{nonlinear_papers1,nonlinear_papers2,guibao2010}. These tensors are particularly
useful because of their high sensitivity to both lattice and
electronic symmetries.  In general, the polarization induced in a
material when it is illuminated by an optical beam can be written as
the Bloembergen expansion \cite{Bloembergen}
\begin{equation}
P_i=\epsilon_0 \chi^{(1)}_{ij} E_i+\epsilon_0 \chi^{(2)}_{ijkl}
E_i E_j +\epsilon_0 \chi^{(3)}_{ijkl} E_j E_k E_l \ldots
\end{equation}
where $E_i$ ($i=x,y,z$) are the components of the electric field,
$\chi^{(1)}_{ij}$ is the linear susceptibility tensor that
determines the linear polarization response of the material, and
$\chi^{(2)}_{ijk}$, $\chi^{(3)}_{ijkl}$...are nonlinear
susceptibility tensors of different ranks responsible for the
nonlinear polarization generated in the medium.

Here we show two  main things. Firstly, that it is possible to
perform complementary symmetry studies of nonlinear materials by
using nonlinear optics processes besides second harmonic
generation, such as spontaneous parametric down-conversion (SPDC).
In SPDC, an intense pump beam with frequency $\omega_p$ interacts
with the atoms or molecules of a second order nonlinear crystal
with nonlinear coefficient $\chi^{(2)}_{ijk}$. In the process, a
flow of paired photons is generated, the signal and idler, with
central frequencies $\omega_s$ and $\omega_i$, such that
$\omega_s+\omega_i=\omega_p$.

Secondly, that instead of illuminating the nonlinear crystal with
paraxial Gaussian beams that propagate along a myriad of different
directions, as it is usually done in similar cases
\cite{nonlinear_papers2}, it is more advantageous to choose as
illuminating beam a non-paraxial optical beam.
The use of a laser source with fast switching of its pointing
direction might be technically cumbersome, and can severely limit
the applicability of the method due to the need to control
mechanical noise. However, the use of an appropriately designed
non-paraxial beam allows to unveil the crystal symmetry group in a
single shot experiment.

Under standard conditions \cite{torres2011}, most SPDC
configurations consist of a Gaussian pump beam that impinges
normally onto the surface of the nonlinear crystal, which is
endowed with a second order nonlinear susceptibility
$\chi_{ijk}^{(2)}$. The cut angle of the crystal is chosen to
guarantee the fulfillment of the phase matching conditions, which
are equivalent to the conservation of energy and momentum of the
photons involved in the SPDC process.  In most
cases, the pump beam  is linearly polarized. Under these
circumstances, the distribution of wave vectors of the signal and
idler photons, for given frequencies of the resulting photon
pairs, is restricted to a single cone (in the case of type I SPDC)
or to two cones (in the case of type II SPDC) with axes
symmetrically arranged relative to the pump beam \cite{kwiat1995}.
These conical distributions are observed for different crystal
symmetries, so they do not bear useful information about the point
group to which the crystal belongs. This is
because in the configuration considered, the nonlinear process
addresses only certain elements of the nonlinear susceptibility
tensor.

We  show that a different configuration  can lead to a spatial
distribution of the photon pairs that directly
 reflects the symmetry group of the crystal. For the sake
of simplicity, we will consider the case of a uniaxial
birefringent crystal with its optics axis parallel to the normal
of its surface. The pump beam that illuminates
the crystal is a vectorial Bessel mode \cite{hbb,YB}, prepared to
guarantee a vectorial extraordinary character inside the nonlinear
media. This can be done by choosing a transverse magnetic (TM)
mode  with its main propagation direction
coincident with the axis of the nonlinear crystal.

 A Bessel beam is formed by the superposition
of plane waves with wave vectors confined in a cone and with a
circular cylindrical symmetry on the angular spectrum. In this
way, the cylindrical symmetry of the photon pairs produced in the
SPDC process is directly broken by the intrinsic symmetry of the
crystal, which manifests on the spatial distribution of the
resulting photon pairs. The axicon angle of the Bessel vectorial
mode is the parameter to be optimized for both the fulfilment of
the phase matching conditions and to obtain a clear visibility of
the characteristic crystallographic pattern built by the photon
pairs.

Note that choosing a Bessel beam as a pump of the nonlinear
process is equivalent to observing the crystal structure
simultaneously for many different angles. In this configuration,
SPDC (or any other nonlinear optical process) is sensible to all
the components of the nonlinear optical susceptibility tensor,
including that along the main direction of propagation of the pump
beam; that is due to the fact that TM modes acquire a significant
component of its electric field along their main direction of
propagation as they depart from the paraxial limit \cite{YB}.

In this paper we consider vectorial monochromatic beams, i.e., beams with an electric field
$\mathbf{\cal{E}} (\mathbf{r},t)=1/2\, \mathbf{E}_{m}(\mathbf{r})
\exp \left( -i \omega t \right) + c.c$ where $\omega$ is the
angular frequency of the beam, $\mathbf{r}$ designates the spatial
location, $t$ is time, $\mathbf{E}_{m}$ is the vectorial
spatially-varying amplitude of the beam and $m$ is related to the
orbital angular momentum of the beam.

In free space, the electric field of a transverse-magnetic (TM)
Bessel beam of order $m$, which is an exact solution of Maxwell's
equations, can be written \cite{hbb} as a sum of plane-waves with
equal amplitude and wavevector confined in a cone around the
$\hat{{\bf z}}$ axis, the main direction of propagation of the beam, with
the so called axicon angle $\varphi_{a}$, i.e.,
\begin{equation}
\mathbf{E}_{m}(\mathbf{r}) = E_0  \int d {\bf p}\,  {\bf e}_k \exp
\left( i k_z z+i {\bf p} \cdot \mathbf{r}_{\perp}+i m \varphi_{\bf p}
\right) \label{bessel_free}
\end{equation}
where $ {\bf e}_k={\hat{{\bf{k}}}}({\hat{{\bf k}}}\cdot{\hat{{\bf z}}}) -{\hat{{\bf z}}}$ is the
polarization of each wave with wavevector ${\bf k}$, the
longitudinal wavenumber is  $k_z=(\omega/c)\,\cos \varphi_a$, and
the transverse wavevector writes ${\bf p}=(\omega/c)\,\sin \varphi_a
\left( \cos \varphi_{\bf p} \hat{{\bf x}}+\sin \varphi_{\bf p} \hat{\bf{y}} \right)$.
$\varphi_{\bf p}$ is the angle between the transverse wavevector ${\bf p}$ and $\hat{{\bf x}}$, and
can take any value between $0$ and $2\pi$. Notice
that even though the polarization of each ${\bf k}$-wave is
perpendicular to ${\bf k}$ \cite{cohen}, the superposition of all ${\bf k}$-waves
yields a beam with a non-zero field component along the main direction
of propagation ($\hat{{\bf z}}$). For paraxial beams ($\varphi_a$ small)
and $m=0$, one obtains the so-called radial modes
\cite{hbb, radial_modes2}.

In experiments, one never  generates such an
ideal beam. However, one can generate vectorial Bessel beams close
to the one described by Eq.~(\ref{bessel_free}) by generating
superpositions of scalar Bessel modes with the proper topological
charge and polarization \cite{afjhrjkv}. Under these experimental
conditions, the wavevectors $\mathbf{k}$ are no longer confined
to a cone of angle $\varphi_a$, with transverse wavevector ${\bf
q}$, but instead the wavevectors spread narrowly around a central
value, ${\bf q}+\Delta {\bf q}$, or equivalently, $\varphi_a +
\Delta \varphi_a$, with $|{\Delta q}|=|{\bf k}| \Delta \varphi_a$.

The Hamiltonian of interaction of SPDC is \cite{louisell1961}
$$ \hat{{\cal H}}(t)=\epsilon_0 \int_{V} dV\int\,d {\bf k}_p\,\int\,
d {\bf k}_s\,\int\,d {\bf k}_i\,$$
\begin{equation}
\label{hamiltonian_spdc}\chi^{(2)}\, {\bf E}_p (t,{\bf r};{\bf k}_p) {\bf E}_s^{-}(t,{\bf r};{\bf k}_s) {\bf
E}_i^{-}(t,{\bf r};{\bf k}_i) + h. c.
\end{equation}
where $V$ is the volume of interaction and $\chi^{(2)}$ is the
nonlinear tensor that characterizes the nonlinear response of the
material.

Let us consider as example a type I ({\em eoo}) SPDC process in an
uniaxial crystal \cite{Dmitriev}. A configuration that will give
us information about the symmetry of the crystal is the one where
the pump beam propagates inside the nonlinear crystal along the
optical axis ($\hat{{\bf c}}=\hat{{\bf z}}$).
The signal and idler waves  propagate as ordinary waves. For the
case of an intense classical extraordinary pump beam that
generates  ordinary signal and  idler photons, the pump beam, and
the electric field operators for signal and idler
photons write \cite{torres2011}
\begin{eqnarray}
 {\bf E}_p^{+}(t,{\bf r};{\bf q}_p)&=&E_0  {\bf e}_p({\bf
q}_p)\,e^{ i m \varphi_{{\bf q}_p}} e^{i k_z^p ({\bf q}_p) z+i{\bf
q}_p \cdot {\bf r}_{\perp}} ,\nonumber \\
 {\bf E}_s^{-}(t,{\bf r};{\bf p})&=&i N_s {\bf e}_s({\bf p})\,
a_s^{\dagger} (k_z^s ({\bf p}), {\bf p})\, e^{ik_z^s ({\bf p})
z+i{\bf
p} \cdot {\bf r}_{\perp}}, \nonumber \\
 {\bf E}_i^{-}(t,{\bf r};{\bf q})&=&i N_i{\bf e}_i({\bf q})\,
a_i^{\dagger}(k_z^i({\bf q}),{\bf q})\, e^{i k_z^i({\bf q}) z
+i {\bf q} \cdot {\bf r}_{\perp}},
\end{eqnarray}
where $N_s$ and $N_i$ are normalization factors,$$k_z^p({\bf q}_p) =\sqrt{\Big(\frac{\omega_p}{c}\Big)^2
\epsilon_{op}-\Big(\frac{\epsilon_{ep}}{\epsilon_{op}}\Big)|{\bf q}_p|^2},$$
$$k_z^s({\bf p})=
\sqrt{\Big(\frac{\omega_s}{c}\Big)^2 \epsilon_{os}
-\vert{\bf p}\vert^2},
\quad k_z^i({\bf q})=
\sqrt{\Big(\frac{\omega_i}{c}\Big)^2 \epsilon_{oi}-\vert{\bf
q}\vert^2},$$
$\epsilon_{op,os,oi}$ are the ordinary relative permittivities
inside the crystal of the pump, signal and idler waves,
$\epsilon_{ep,es,ei}$ are the corresponding extraordinary ones,
and
$$ {\bf e}_p ({\bf  q}_p)=\Big[ \frac{c^2}{\omega_p^2
\epsilon_{op}}
{\bf k}_p ({\bf k}_p \cdot \hat{\mathbf{z}}) -\hat{{\mathbf z}} \Big],$$
\begin{equation}
 {\bf e}_s({\bf  p})={\bf k}_s \times  \hat{{\bf c}},\quad
 {\bf e}_i({\bf q})= {\bf k}_i \times  \hat {{\bf c}}
\end{equation}
$a_s^{\dagger}(k_z^s({\bf p}),{\bf p})$ and
$a_i^{\dagger}(k_z^i({\bf p}),{\bf q})$ are creation operators for
signal and idler photons, with momentum $k_z^s \hat{z}+{\bf p}$
and $k_z^i \hat{z}+{\bf q}$, respectively.

Since in most situations the nonlinear interaction is weak, we can
obtain an accurate quantum description by calculating the
first-order solution of the Schr\"{o}dinger equation, i.e.,
\begin{equation}
\label{schrodinger2} |\Psi(t) \rangle=\ket{\text{vac}}-\left(
\frac{i}{\hbar}\right)\int_{-\infty}^{t} dt^{\prime} {\cal
H}(t^{\prime})|\text{vac}\rangle_s|\text{vac}\rangle_i
\end{equation}
The interaction Hamiltonian is effectively zero when the classical
beams that pump the nonlinear process are zero, so that, the time
of integration can be extended to $t=\infty$.

The quantum state of the down-converted photons at the output face
of the nonlinear crystal, neglecting for the sake of simplicity
the contribution from the vacuum term can be written as
\begin{equation}
|\Psi\rangle \sim \int d{\bf p} d{\bf q} F({\bf p}, {\bf q})
a_s^{\dagger} (k_z^s,{\bf p}) a_i^{\dagger} (k_z^i,{\bf q})
|\text{vac} \rangle
\end{equation}
where
\begin{eqnarray}
& & F({\bf p}, {\bf q}) =\sum_{l,m,n}  \chi^{(2)}_{lmn} \left[
{\bf e}_p({\bf p}+{\bf q})
\right]_l \left[ {\bf e}_s({\bf p}) \right]_m \left[ {\bf e}_i({\bf q}) \right]_n \nonumber \\
& & \times   \text{sinc} \frac{\Delta k_z({\bf p},{\bf q})
L}{2} \nonumber \\
& & \times \exp \left[ i \frac{k_z^p({\bf p}+{\bf q})+k_z^s({\bf
p})+k_z^i({\bf q})}{2}L\right]
\end{eqnarray}
and $\Delta k_z ({\bf p},{\bf q})=k_z^p({\bf p}+{\bf
q})-k_z^s({\bf p})-k_z^i({\bf q})$.

It is important to remind here the crucial role that the tensorial
character of $\chi^{(2)}$ plays in the SPDC configuration
considered. Most experiments that make use of SPDC, due to the paraxial
character of the pump, can be described using an effective nonlinear index \cite{Dmitriev} that writes
$\chi^{(2)}_{eff}=\left( \hat{{\bf e}}_p \right)_{\bot} \left[
\chi^{(2)} \right] \left( \hat{{\bf e}}_s \right)_{\bot} \left( \hat{{\bf e}}_i
\right)_{\bot}$, where $\hat{{\bf e}}_{p,s,i}$ are the
linear polarizations of the pump, idler and signal photons, respectively.  The flux of
down-converted photons depends on the magnitude of the effective
index. The vectorial structure of the
electromagnetic field  defines its polarization and,
consequently, affects its total angular momentum. Thus, non paraxial pump
beams are expected to yield interesting results determined by the
structure of $\chi^{(2)}_{lmn}$ and its relation to the angular
momentum content of the down-converted photons \cite{rj}.

\begin{figure}[t!]
\begin{tabular}{@{}c@{}c}
      \subfigure[]{\includegraphics[width=0.4\textwidth,trim=35mm 30mm 20mm 85mm, clip=true]{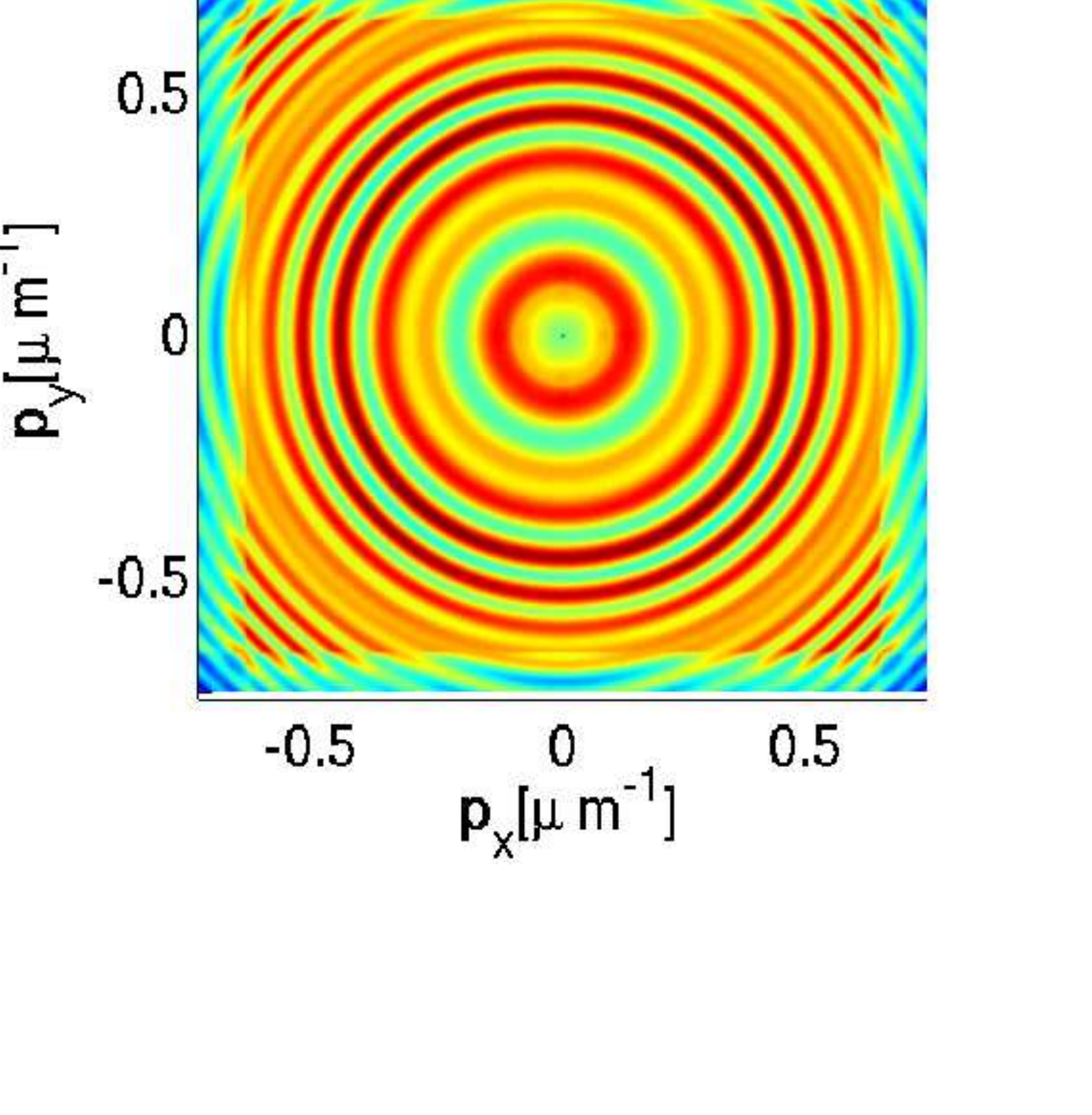}}&
      \subfigure[]{\includegraphics[width=0.4\textwidth,trim=35mm 30mm 20mm 85mm,clip=true]{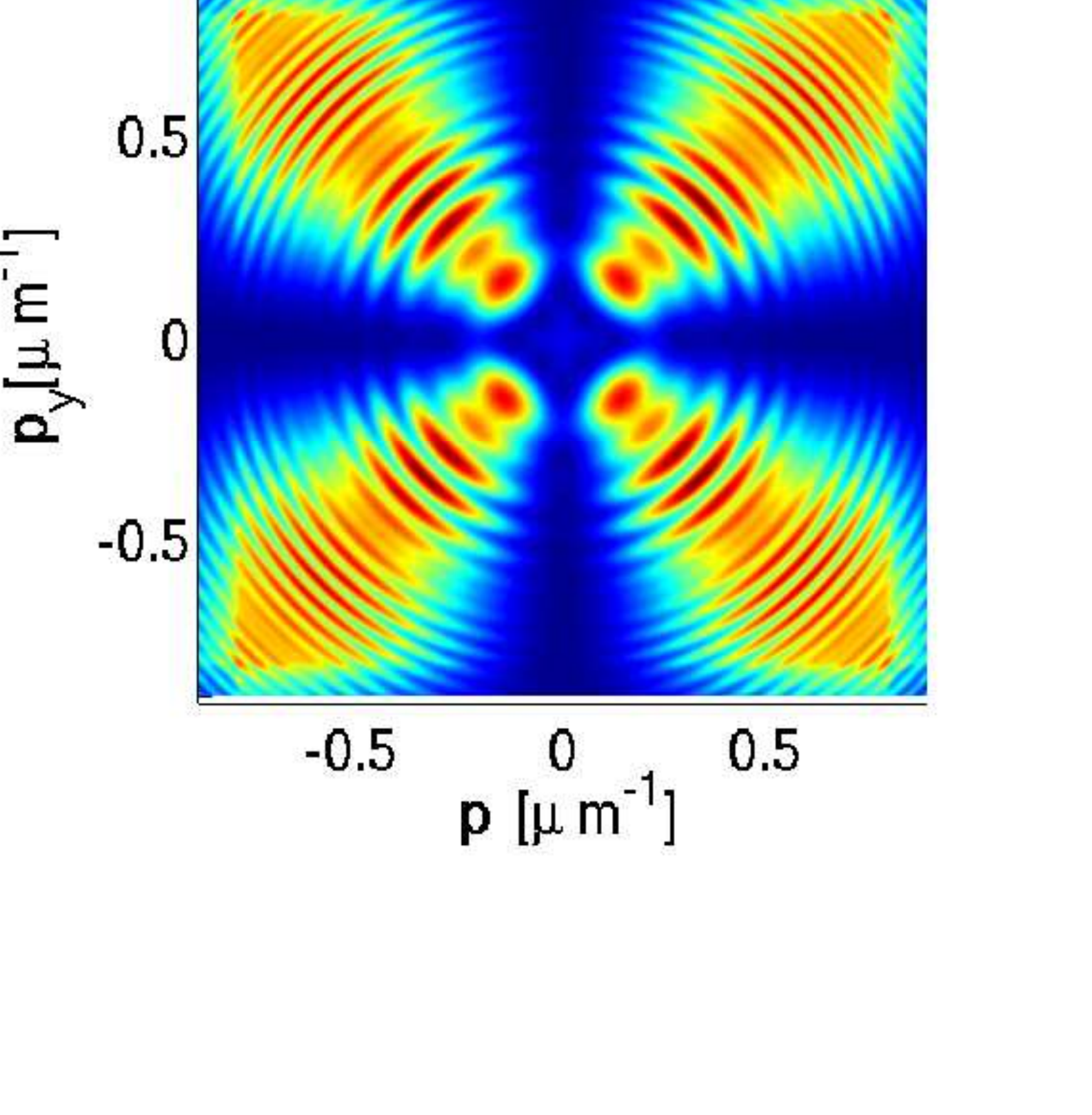}}\\

      \subfigure[]{\includegraphics[width=0.4\textwidth,trim=25mm 30mm 20mm 85mm,clip=true]{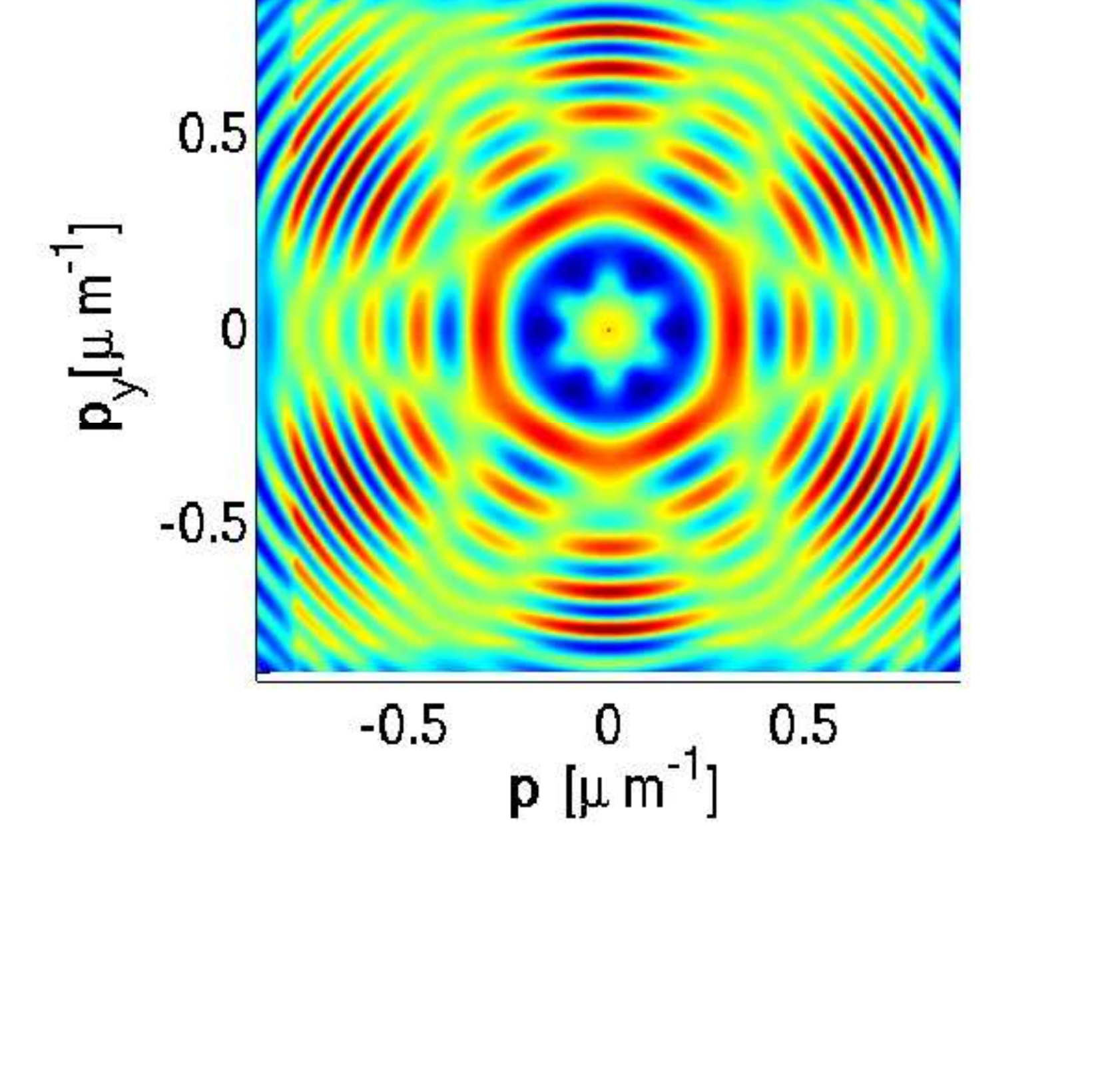}}&
      \subfigure[]{\includegraphics[width=0.4\textwidth,trim=25mm 30mm 20mm 85mm,clip=true]{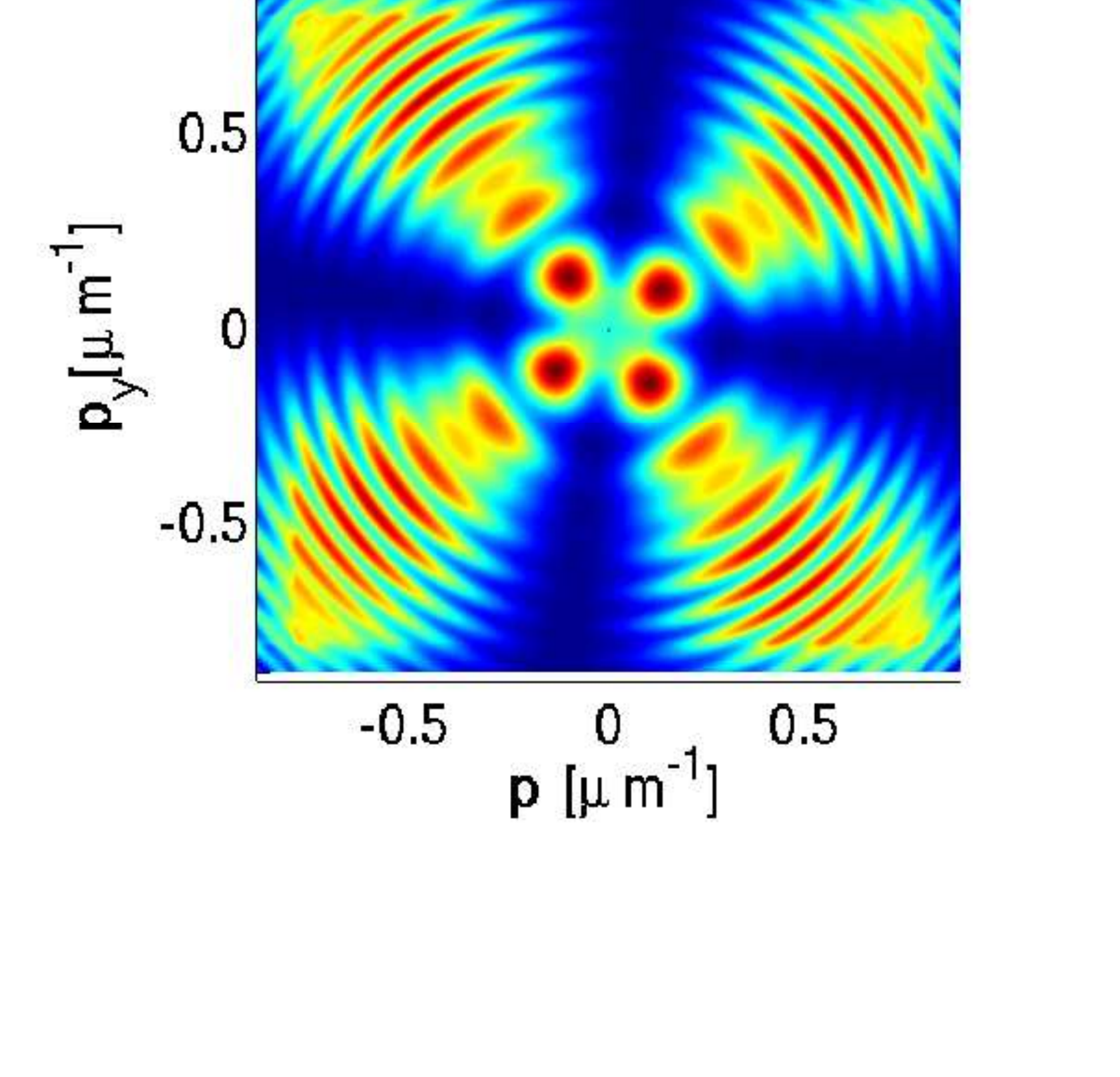}}\\

      \subfigure[]{\includegraphics[width=0.4\textwidth,trim=30mm 30mm 10mm 85mm,clip=true]{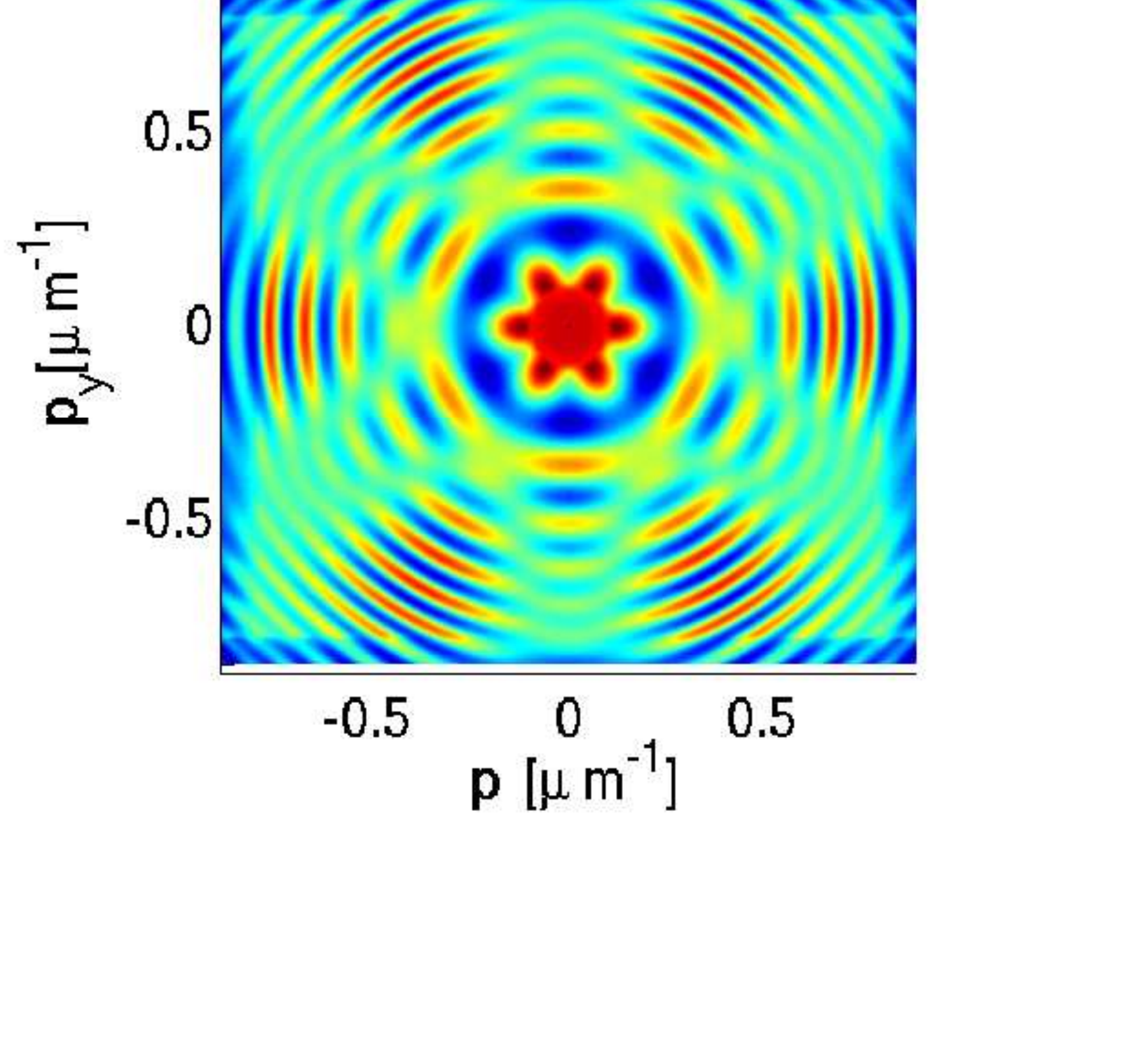}}&
      \subfigure[]{\includegraphics[width=0.41\textwidth,trim=25mm 30mm 10mm 85mm,clip=true]{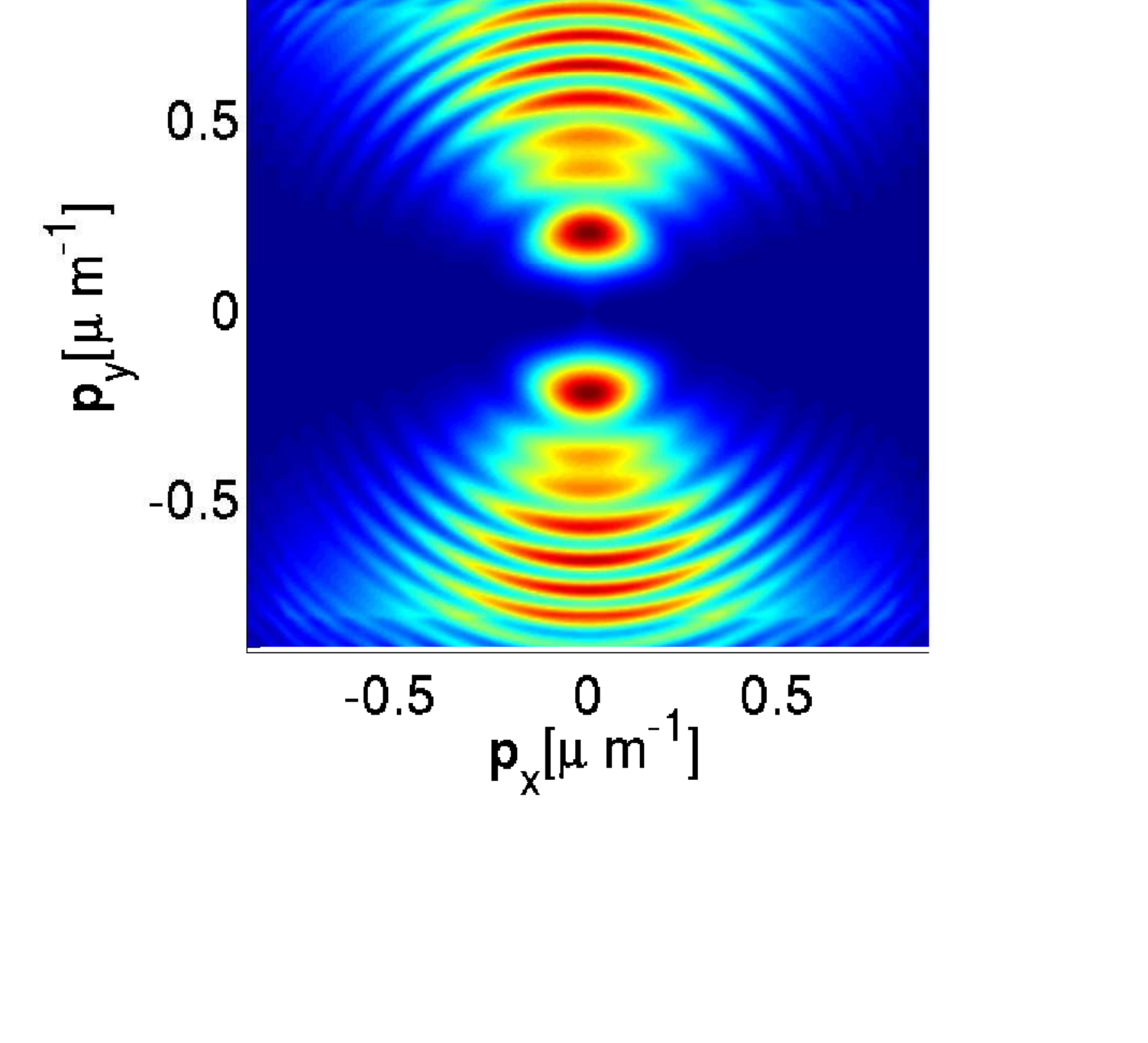}}\\
\end{tabular}
\caption{Angular dependence of the flux rate of
 signal photons detected [{\em singles
detection} given by Eq. (\ref{singles_detection})] for type I
degenerate SPDC ($\lambda_s=\lambda_i=2\lambda_p$). The wavelength
of the incident TM Bessel mode is $\lambda_p=407$ nm. The angle of
the cone that characterizes the illuminating pump Bessel beam
is $\sin \varphi_a = 0.0097 $. We also consider a
spread of directions $\Delta \varphi_a \sim 0.0004$ rad. The
crystal length is taken as $L=1$ mm.
(a) CdSe: point group 6mm; (b) KDP: point group $\bar{42}$m;
(c) GaSe: point group
$\bar{62}$m; (d) HgGa$_2$S$_4$: point group $\bar{4}$; (e) HgS:
point group 32; (f) LiNbO$_3$:point group $3$m}
\end{figure}

In general, $F({\bf p}, {\bf q})$ describes an entangled state in
the polarization and transverse wavevector degrees of freedom.
This entanglement takes place both when considering the signal and
idler photons as the two subsystems of the whole system, and when
considering the two degrees of freedom, polarizations and
transverse wavevectors, as the corresponding subsystems.

The flux rate  $R_{si}({\bf p},{\bf q})$ of detections of a signal
photon with transverse wavevector ${\bf p}$ in coincidence with an
idler photon with transverse wavevector ${\bf q}$ ({\em
coincidence detection}), or equivalently, a signal photon that
propagates inside the crystal along the direction
($\theta_s$,$\varphi_s$) with $\theta_s=\tan^{-1} |{\bf
p}|/\sqrt{(\omega_s/c)^2 \epsilon_0-|{\bf p}|^2}$ and
$\varphi_s=\cos^{-1} p_x/|{\bf p}|$ and an idler photon that
propagates along the direction ($\theta_i$,$\varphi_i$) with
$\theta_i=\tan^{-1} |{\bf q}|/\sqrt{(\omega_i/c)^2
\epsilon_0-|{\bf q}|^2}$ and $\varphi_i=\cos^{-1} q_x /|{\bf q}|$
is
\begin{equation}
R({\bf p},{\bf q})=|F({\bf p}, {\bf q})|^2
\label{coincidence_detection}
\end{equation}
while the flux rate $R_s({\bf p})$ of signal photons detected with
transverse wavevector ${\bf p}$ ({\em singles detection}) writes
\begin{equation}
R_s({\bf p})= \int d{\bf q} R({\bf p},{\bf q})=\int d{\bf
q}|\,F({\bf p}, {\bf q})|^2. \label{singles_detection}
\end{equation}
We illustrate these results, and how they help
to determine the symmetry group of the crystal, by showing some
examples. In Fig.~1, we exemplify the angular dependence of the flux
rate of signal photons (singles detections as given by Eq.~(\ref{singles_detection}))
for a variety of nonlinear crystals
with different crystallographic symmetries. The numerical
simulations consider the adequate Sellmeier equations and the
crystal birefringent properties. For small values of the
transverse wave vector of the pump beam ($|{\bf q}_p| \le 0.01
\mu$m$^{-1}$ for most of the crystals
considered here, which corresponds to an angle $\varphi_a \ll
0.0097$), the angular spectrum is always formed by a set of
concentric rings similar to those observed  for
$|{\bf p}_s| \le 0.5\mu$m$^{-1}$ in Fig.~1(a). As $|{\bf q}_p|$
increases the down-converted photons are emitted in a wider region
in transverse wavevector space, and the symmetry of the crystal
becomes more evident. Properties such as the radius of the
concentric rings, the values of $|{\bf q}_p|$ at which the
visibility of the symmetry of the crystal is significant, and the
maximum value of $R_s$, could be used to determine some
characteristics of the linear susceptibility tensor and the
second- order nonlinear susceptibility tensors. For instance, KDP
and CsH$_2$AsO$_3$ belong to the same point group, $\bar{42}$ m,
but they have slightly different values of $\epsilon_o$,
$\epsilon_e$ and the relevant components of $\chi_{lmn}^{(2)}$.
This manifests in general common properties between the angular
spectrum of both crystals, but there are still measurable
differences for small values of ${\bf q}_s$ due to a higher
sensitivity of CsH$_2$AsO$_3$ to the electric field component
along the $z$-axis.

\begin{figure}[t!]
\begin{tabular}{@{}c@{}c}
\subfigure[]{\label{f:Rcas:GaSe:0p15}\includegraphics[width=0.5\textwidth,trim=35mm 30mm 20mm 85mm, clip=true]{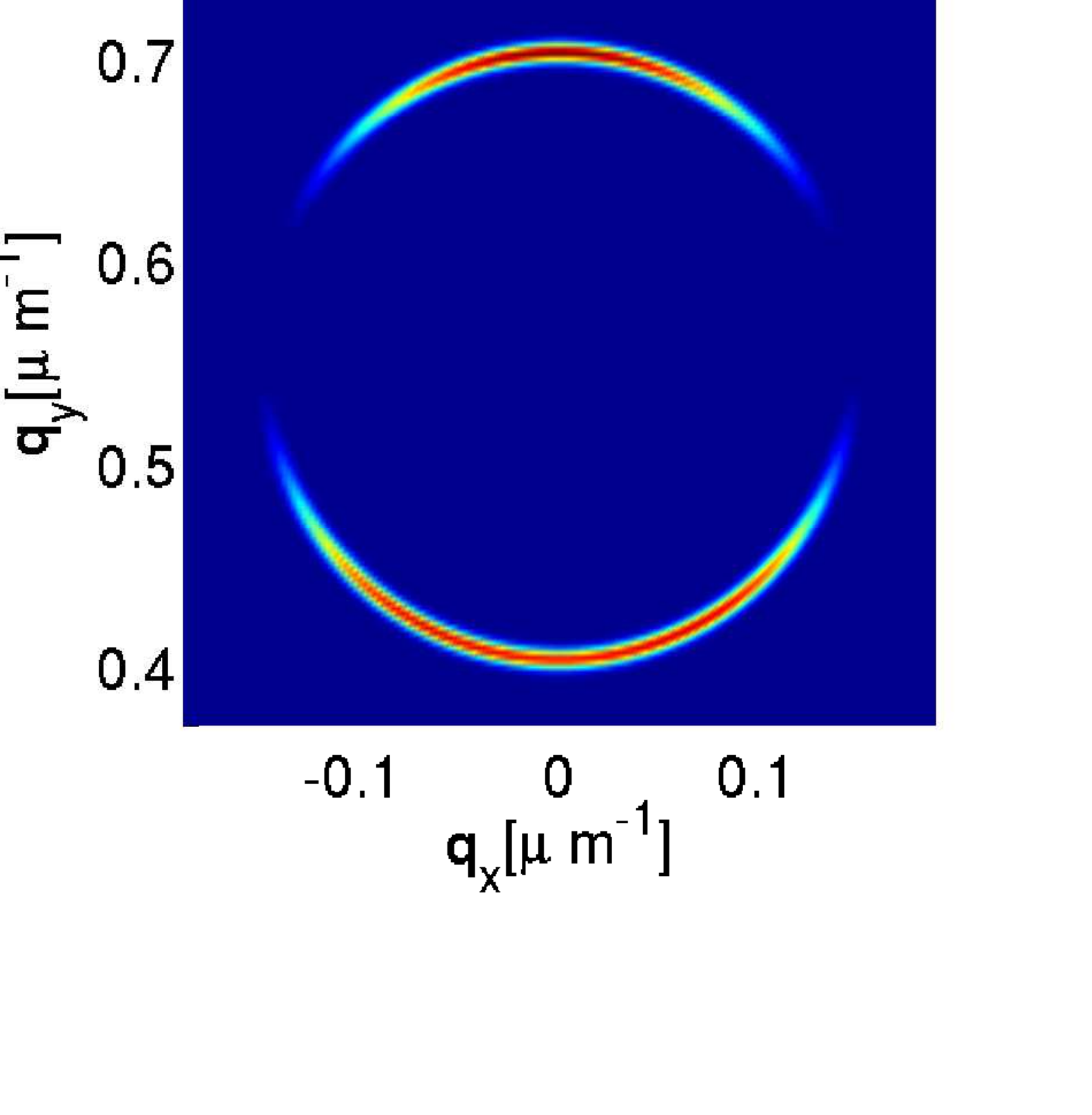}}&
\subfigure[]{\label{f:Rcas:BBO:0p15}\includegraphics[width=0.5\textwidth,trim=35mm 30mm 20mm 85mm, clip=true]{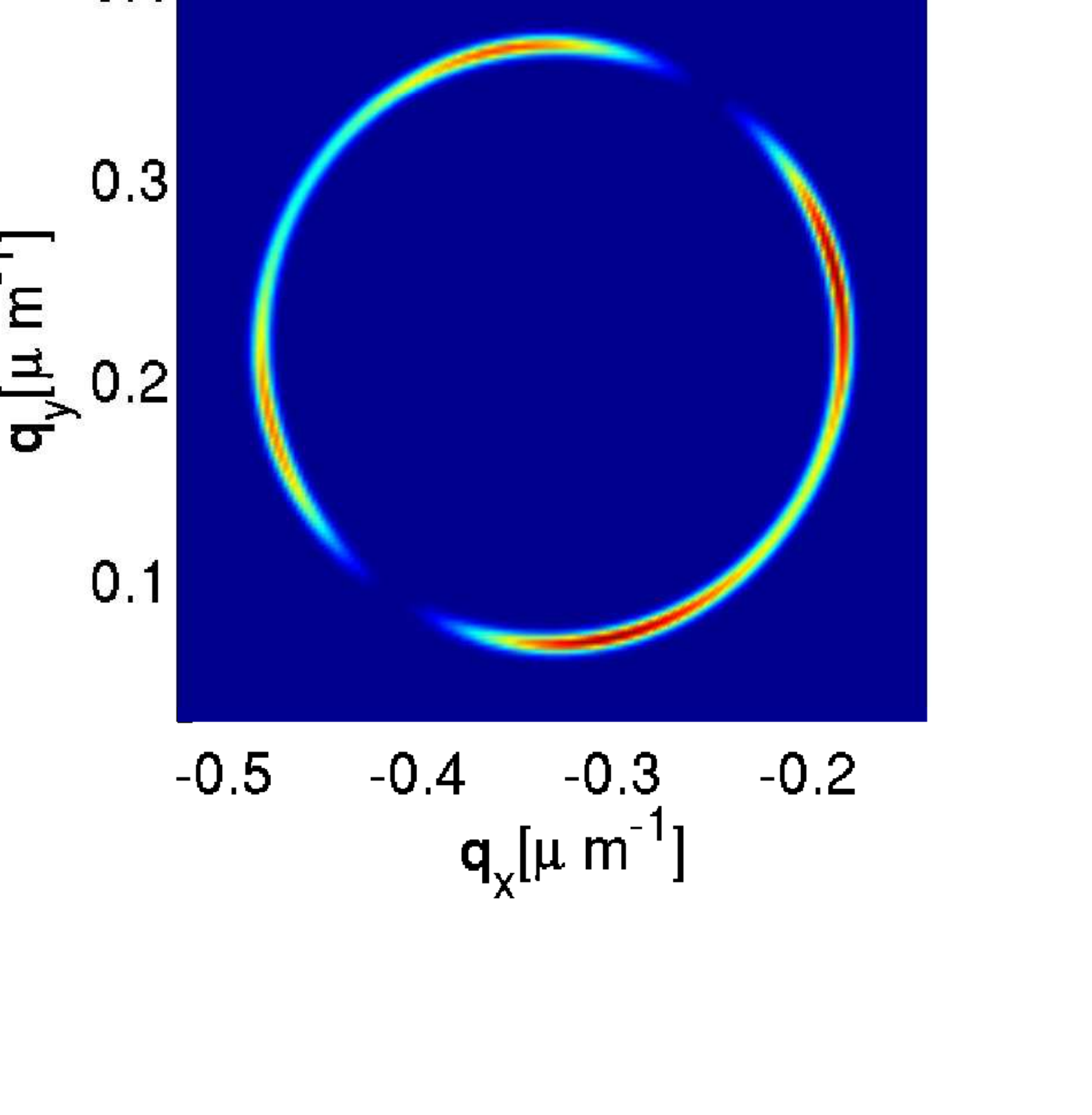}}\\
\subfigure[]{\label{f:Rcas:LiNbO3:0p15}\includegraphics[width=0.5\textwidth,trim=35mm 30mm 20mm 85mm, clip=true]{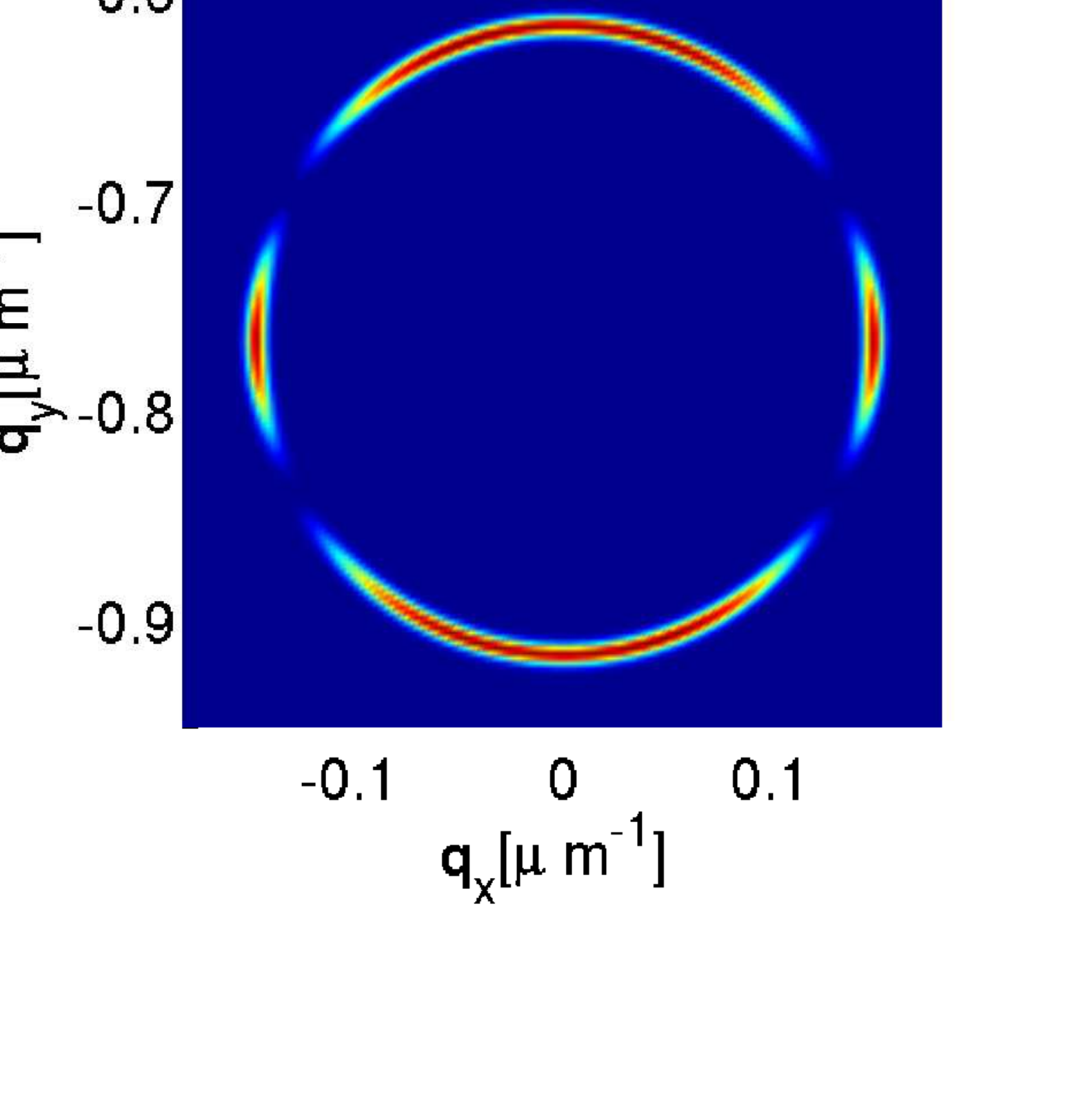}}&
\subfigure[]{\label{f:Rcas:KDP:0p15}\includegraphics[width=0.5\textwidth,trim=35mm 30mm 20mm 85mm, clip=true]{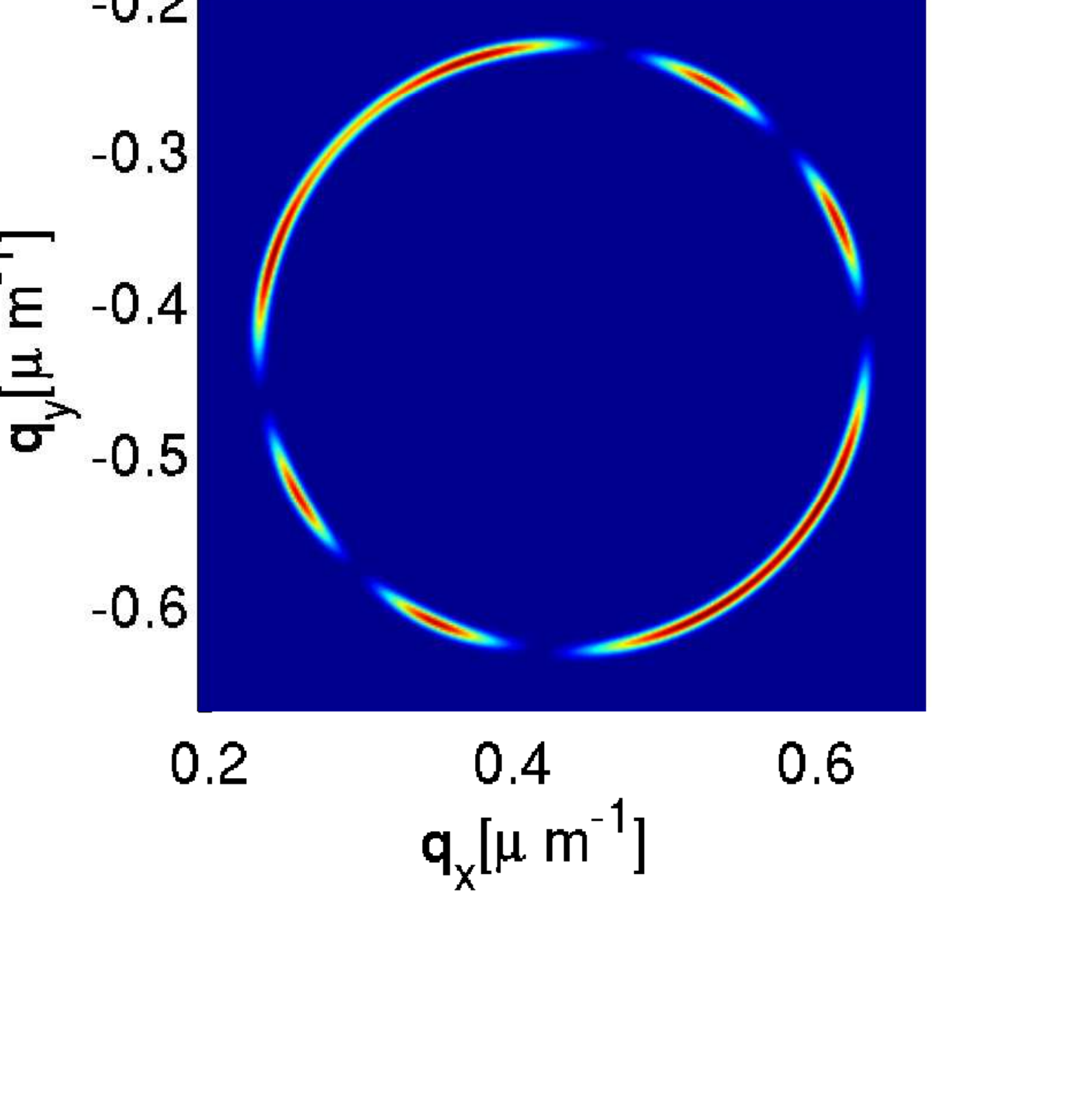}}
\end{tabular}
\caption{Flux rate of idler photons detections in coincidence with
the signal photons with the transverse wavevector ${\bf p}$ that
maximizes $R_{s}$. We consider type I degenerate SPDC in a
nonlinear crystal of length $L$ = 1mm: (a) GaSe; (b)BBO;
(c)LiNbO$_3$ and (d) KDP. The incident beam is a Bessel TM mode
with wavelength $\lambda$ = 407nm, $|{\bf q}_p| = 0.15
\mu$m$^{-1}$ for (a-c) and $|{\bf q}_p| = 0.2 \mu$m$^{-1}$ for
(d). Again, we consider an spread of values of ${\bf q}_p$ of
$\Delta {\bf q}_p = 1/150 \mu$m.}
\end{figure}

The coincidence detections, given by Eq.~(\ref{coincidence_detection})
also provide useful information
about the symmetry group of the medium under investigation. A
direct calculation shows that for $|{\bf q}_p|<<\omega_p/c$, the
detection of a signal photon with transverse wavenumber ${\bf p}$
is accompanied by the detection of an idler photon that is
confined to a ring with radius $|{\bf q}| \sim |{\bf q}_p|
\epsilon_o/\epsilon_e$ around ${\bf q}=-{\bf p}$. As ${\bf q}_p$
increases, the symmetry of the crystal becomes more visible in the
angular distribution of idler photons, the flux rate becomes
inhomogeneous along the ring with a structure that depends on the
symmetry of the crystal, and which is compatible with the
conditional generation of stationary Bessel photons. That is,
photons with an angular spectra resulting from superpositions of
$m$ and $-m$ Bessel modes. In Fig.~2, this effect is illustrated
for various types of crystals.

Summarizing, we have shown that, under the adequate experimental
set up, the correlations of the photons emitted in an SPDC
process, contain crystallographic information that can be accessed
by measurements of the SPDC angular spectra and the conditional
angular spectra. It is expected that a complete characterization
of the twin photons correlations as a function of the properties
of the structured pump beam could be used to obtain precise
measurements of the first and second order electric susceptibility
tensors. Notice that the benefits arising from structured pump
beams for nonlinear crystallography should not be restricted to
the SPDC process. In fact, nonlinear harmonic generation
crystallography \cite{nonlinear_papers1} with Bessel beams could be an
alternative to rotational anisotropy measurements \cite{nonlinear_papers2},
since using Bessel modes as a pump is equivalent to  simultaneous
measurements along different directions.

\begin{acknowledgments} JPT acknowledges support from the
program Severo Ochoa of the Government of Spain, the Fundacio
Privada Cellex Barcelona and the program ICREA ACADEMIA from the
Generalitat de  Catalunya. RJ acknowledges partial support
from the  grants CONACyT  CB-166961 and LN-232652.
\end{acknowledgments}


\begin{thebibliography}{99}

\bibitem{sands} D. E. Sands, ``Introduction to crystallography'',
Dover publications, New York, 1969.

\bibitem{neumann_principle} R. E. Newnham,``Properties of materials: anisotropy, symmetry, structure",
 Oxford University Press, New York, 2005.

\bibitem{x-ray} J. A. K. Howard, and M. R. Probert, ``Cutting-edge techniques used for the structural investigation of single crystals",
 {\it Science} {\bf 343}, 1098 (2014).

\bibitem{neutron-waves} T. F. Koetzle and G. J. McIntyre, ``Characterization of materials",
 Wiley  Online Library, 2012; G. L. Squires, ``Introduction to the theory of thermal neutron scattering",
Cambridge University Press, 1978.

\bibitem{electron-waves}X. D. Zou, S. Hovm\"oller, and P. Oleynikov, ``Electron crystallography: electron microscopy and electron diffraction", Oxford University Press, 2011.

\bibitem{nonlinear_papers1}  H. W. K. Tom, T. F. Heinz, and Y. R. Shen, ``Second-harmonic reflection from Silicon surfaces and its relation to structural symmetry", {\it Phys. Rev.
Lett.} {\bf 51}, 1983 (1983).

\bibitem{nonlinear_papers2}D. H. Torchinsky, Hao Chu, T. Qi, G. Cao, and D. Hsieh, ``A low temperature nonlinear optical rotational anisotropy spectrometer for the determination of crystallographic and electronic symmetries",
{\it Rev Sci. Inst.} {\bf 85}, 083102 (2014).

\bibitem{guibao2010} G. Xu, G. Sun, Y. J. Ding, I. B. Zotova,
K. C. Mandal, A. Mertiri, G. Pabst, and N. Fernelius, ``Investigation of symmetries of second-order nonlinear susceptibility tensor of GaSe crystals in THz domain", {\it Opt. Comm.} \textbf{284}, 2027 (2011).

\bibitem{Bloembergen} N. Bloembergen, ``Nonlinear optics", World Scientific Pub., 4th edition, 1996.


\bibitem{torres2011} J. P. Torres, K. Banaszek, and I. A. Walmsley, ``Engineering nonlinear optic sources of photonic entanglement",
{\it Progress in Optics} \textbf{56}, 227 (2001).

\bibitem{kwiat1995} P. G. Kwiat, K. Mattle, H. Weinfurter, A. Zeilinger, A. V.
Sergienko, and Y. Shih, ``New high-intensity source of polarization-entangled photon pairs", {\it Phys. Rev. Lett.} \textbf{75}, 4337 (1995).



\bibitem{hbb} R. Horak, Z. Bouchal, and J. Bajer, ``Nondiffracting stationary electromagnetic field", {\it Opt. Commun.} {\bf 133},
315 (1997).

\bibitem{YB} K. S. Youngworth and T. G. Brown, ``Focusing of high numerical aperture cylindrical-vector beams",
{\it Opt. Express} {\bf 7}, 77 (2000).

\bibitem{cohen} C. Cohen Tannoudji, J. Dupont-Roc and G. Grynberg, ``Photons and atoms: introduction to quantum electrodynamics",
John Wiley and sons, New York, 1997.


\bibitem{radial_modes2} A. Dudley, Y. Li, T. Mhlanga, M. Escuti, and A. Forbes, ``Generating and measuring nondiffracting vector Bessel
beams", {\it Opt. Lett.} {\bf 38}, 3429 (2013).


\bibitem{afjhrjkv} A. Flores-P\'erez, J. Hern\'andez-Hern\'andez, R. J\'auregui, and K. Volke-Sep\'ulveda, ``Experimental generation and analysis of first-order
TE and TM Bessel modes in free space", {\it Opt. Lett.} \textbf{31}, 1732 (2006).

\bibitem[Louisell (1961)]{louisell1961} W. H. Louisell, A. Yariv and A. E. Siegman, ``Quantum fluctuations and noise in parametric processes. I",{\it Phys. Rev.} \textbf{124}, 1646 (1961).

\bibitem{Dmitriev} V. G. Dimitriev, G. G. Gurzadyan, and D. N. Nikogosyan, `` Handbook of nonlinear crystals",
{\it Springer Series on Optical Sciences} \textbf{64} (1999).

\bibitem{rj} R. J\'auregui, ``Spontaneous parametric down conversion of vectorial beams: helicity effects on the orbital angular momentum of the photon pairs", invited comment {\it Phys. Scripta} \textbf{90}, 068012 (2015).

\bibitem{SHRJ} S. Hacyan and R. J\'auregui, ``Evolution of optical phase and polarization vortices in birefringent media",{\it J. Opt. A: Pure Appl. Opt.} {\bf 11}, 085204 (2009).

\bibitem{FED} T. A. Fadeyeva, V. G. Shvedov, Y. V. Izdebskaya, A. V. Volyar, E. Brasselet, D. N. Neshev, A. S. Desyatnikov,
W. Krolikowski, and Y. S. Kivshar,``Spatially engineered polarization states and optical vortices in uniaxial crystals",
 {\it Opt. Express} \textbf{18}, 10848 (2010).


\end{thebibliography}
\end{document}